\documentclass[11pt]{article}
\usepackage[margin=1in]{geometry}
\usepackage{amsmath,amssymb}
\DeclareMathOperator*{\argmax}{arg\,max}
\usepackage{graphicx}
\usepackage{booktabs}
\usepackage{float}
\usepackage{hyperref}
\usepackage{xcolor}

\title{\textbf{Red-Blue Hamiltonian Tomography via the Operator Lightcone Wavemap}}

\author{Paolo D'Alberto\thanks{Developed in collaboration with Claude (Anthropic).}\\
  \texttt{paolo.dalberto@amd.com}\\
  Advanced Micro Devices, Inc.
}
\date{}

\begin{document}
\maketitle

\begin{abstract}
We introduce a model-free method for Hamiltonian tomography based on the
operator spreading wavefront. Given two tensor network simulations of the same quantum lattice
(one with the nominal Hamiltonian and one with one or more perturbed bonds), the signed
cross-entropy of the non-identity Pauli weight field
$\mathcal{L}(v,t) = -\ln(n_{\rm pert}/n_{\rm base})$ identifies each perturbed
bond by its causal signature: a red-blue sign flip at the two endpoints, visible
from the first wavefront crossing.
Combining measurements from $p$ strategically placed observation centers via an
ensemble log-likelihood sum, a single broken bond is identified with an 82-fold
margin over the next candidate on a $7\times7$ lattice. Extending to realistic
coupling degradation ($\delta J = -0.20$) and up to 8 simultaneous faults, the
integrated log-ratio $\Lambda(v) = \ln(\mathcal{B}(v)/\mathcal{N}(v))$ achieves
a per-bond detection rate above 90\% in a single seismic pass using only 6
simulations. An iterative re-scan strategy progressively resolves multi-bond cases
as the signal clears with each recalibration pass.
The method requires no model assumptions and no knowledge of the perturbation
type or magnitude; the detection is deliberately blind.
\end{abstract}

\section{Introduction}

Identifying miscalibrated or anomalous couplings in a quantum processor is
a central challenge in quantum hardware characterization. Standard approaches
require explicit Hamiltonian learning: fitting a parameterized coupling model
to measurement data. This scales poorly with system size and depends on model
assumptions about the noise and interaction structure.

We propose a different approach based on a clear separation of roles.
The \emph{simulator} knows the nominal Hamiltonian and produces two Pauli
weight fields: $n_{\rm base}(v,t)$ from the nominal evolution, and
$n_{\rm pert}(v,t)$ from an evolution where one bond has been silently
modified. The \emph{detection algorithm} receives only these
two arrays of numbers and must identify the modified bond from their ratio
alone, with no knowledge of the Hamiltonian, the perturbation type, or
its magnitude. The detection is deliberately blind.

A perturbed bond creates a causal signature in the wavefront: a sign flip
in the cross-entropy field $\mathcal{L}(v) = -\ln(n_{\rm pert}/n_{\rm base})$
between its two endpoints that propagates no further. The broken bond is
exactly where the wavefront sign changes, identifiable from the observables
alone.

The method is inspired by seismic tomography: place sensors (observation
centers) at strategic positions, emit a wave (initialize a local operator),
and read the arrival pattern. The perturbed bond is the anomaly in the medium.

The red-blue color scheme is a deliberate reference to the red-blue pebble
game of Hong and Kung~\cite{hong1981pebble}, which characterizes I/O
complexity as a bottleneck in computational graphs. Here the roles are
analogous: red sites (upstream, wave blocked) and blue sites (downstream,
wave starved) identify the bond that limits operator propagation, exactly
as red and blue pebbles identify the I/O bottleneck in a computation.

\section{Setup}

We simulate the kicked isotropic Heisenberg model on a $7\times7$ rectangular
lattice (49 sites, 84 bonds) using a GPU-accelerated belief propagation tensor
network simulator with Trotter step $\varepsilon=0.10$, bond dimension
$\chi=50$, and truncation cutoff $\delta=10^{-3}$.
The non-identity Pauli weight at site $v$ and cycle $t$ is:
\begin{equation}
    n(v,t) = \sum_{P:\,P_v \neq I} c_P(t)^2
\end{equation}
where $O_c(t) = \sum_P c_P(t) P$ is the evolved operator initialized at
center $c$. By normalization $\sum_P c_P^2 = 1$, $n(v,t) \in [0,1]$ is
a probability.

To simulate miscalibrated bonds, we weaken their couplings by $\delta J$,
with each bond potentially taking a different value. In the experiments below
we use $\delta J = -0.20$ ($J = 0.80$, a 20\% degradation) as a representative
choice; the detection method does not depend on knowing $\delta J$ in advance
and works for any coupling deviation. The Pauli transfer matrix entries scale as
$\cos(J\varepsilon)$, attenuating operator propagation without blocking it entirely.
For the multi-bond experiments, $k$ bonds are selected uniformly at random and perturbed simultaneously.

We place $p=6$ observation centers using a greedy geodesic bond cover: the
minimum set of sites such that every bond is within $d_{\rm max}=2$ hops of
at least one center. For the $7\times7$ lattice this gives centers at sites
$\{6, 10, 25, 30, 32, 34\}$, covering all 84 bonds.

\section{Signal Vocabulary and Measurement Definitions}
\label{sec:vocab}

All definitions are expressed in terms of two observables per site per
cycle: $n_{\rm base}(v,t)$ and $n_{\rm pert}(v,t)$, the non-identity
Pauli weights under the nominal and perturbed Hamiltonian
respectively.  The BFS frontier crossing time $t_0(v) = \min\{t :
n_{\rm base}(v,t) > \varepsilon_{\rm th}\}$ defines a site-specific
reference time; it marks the physical arrival of the wavefront at $v$,
set by the Lieb-Robinson lightcone, not by the numerical cutoff. 

At each site $v$ in the 3-cycle window $[t_0(v) \pm 1]$ we observe one of three
frontier patterns. A site is \textbf{White} $[\mathsf{W}]$ when
$n_{\rm base}(v, t_0) \leq \varepsilon_{\rm th}$: the baseline wave has not yet
arrived and no information is available. A site is \textbf{Blue} $[\mathsf{B}]$
when $n_{\rm pert}(v, t_0) > n_{\rm base}(v, t_0) > \varepsilon_{\rm th}$: the
perturbed wave accumulates at $v$ because drainage is blocked by a broken bond
downstream; the site is a \emph{wall face}. A site is \textbf{Red} $[\mathsf{R}]$
when $n_{\rm pert}(v, t_0) < n_{\rm base}(v, t_0)$ with
$n_{\rm base}(v, t_0) > \varepsilon_{\rm th}$: the perturbed wave is attenuated
because a broken bond on the path from center blocked transmission; the site is
in the \emph{shadow}.

For a bond $(u, v)$ with $u$ on the center-side and $v$ on the shadow-side,
these patterns combine into three observable signatures. When both endpoints
are white $[\mathsf{W,W,W}]$ the bond has not yet been reached or is healthy.
When the wall face is detected but the shadow is not yet visible $[\mathsf{W,B,W}]$,
one more cycle is needed. The \textbf{broken bond signature} $[\mathsf{W,B,R}]$
is the primary target: $u$ appears blue at cycle $t_0$ and $v$ appears red at
cycle $t_0+1$, identifying the bond between them as anomalous.
This pattern rests on a single assumption: both waves reach $u$ at the same
frontier time $t_0$, so any amplitude difference at $v$ is attributable to the
bond, not to a timing offset. The temporal order (blue before red by one cycle)
is then universal, following from the BFS causal structure: $u$ is reached
before $v$, so the wall accumulates before the shadow appears.

The signed cross-entropy evaluated in the frontier window is:
\begin{equation}
    \mathcal{L}(v_0,\, [t_0 \pm 1]) = -\ln \frac{n_{\rm pert}(v_0,\, t_0)}
                                                  {n_{\rm base}(v_0,\, t_0)}
\end{equation}
This is a \emph{cross-correlation fixed in space}: both waves evaluated at the same
site $v_0$ at the same frontier moment $t_0$, with the window $[t_0 \pm 1]$
providing context (white before, red after).
Positive $\mathcal{L}$ identifies red (shadow), negative identifies blue (wall face).
The broken bond is the edge between consecutive blue and red sites in the BFS ordering.

A complementary observable measures the wave \emph{through} a bond rather than \emph{at} a site.
Consider the bond $(v_0, u_0)$ in the BFS path from the center.
The baseline reaches $v_0$ at frontier time $t_0$ and $u_0$ at $t_1 = t_0 + 1$.
The \emph{baseline transfer ratio} is:
\begin{equation}
    T_{\rm base}(v_0 \to u_0) = \frac{n_{\rm base}(u_0,\, t_1)}{n_{\rm base}(v_0,\, t_0)}
\end{equation}
This measures how much of the wave at $v_0$ propagates through the bond to $u_0$ in one BFS step,
directly encoding the gate's coupling strength $J$.
For the perturbed run (with potentially delayed frontier times $t_0^p$, $t_1^p$):
\begin{equation}
    T_{\rm pert}(v_0 \to u_0) = \frac{n_{\rm pert}(u_0,\, t_1^p)}{n_{\rm pert}(v_0,\, t_0^p)}
\end{equation}
The ratio $T_{\rm base} / T_{\rm pert} > 1$ signals a broken bond: less weight
crosses the bond in the perturbed run.

Two important caveats apply. First, there is a path distortion. The
perturbed wave arriving at $v_0$ has already traversed all upstream
bonds. If any of those are also broken, $n_{\rm pert}(v_0, t_0^p)$ is
already weakened, not by bond $(v_0, u_0)$ but by upstream damage.
The ratio $T_{\rm base}/T_{\rm pert}$ then conflates the upstream
distortion with the target bond signal. The ratio is not a pure
measurement of bond $(v_0, u_0)$. Second, there is a measurable
Sensitivity. problem. The perturbed weight $n_{\rm pert}(v_0, t_0^p)$
may be close to $\varepsilon_{\rm th}$, the threshold, making the
ratio numerically unreliable.  The local cross-correlation
$\mathcal{L}(v_0, [t_0 \pm 1])$ avoids both caveats by using the
baseline as the reference; it measures the \emph{relative} change at
the frontier moment, not the absolute transfer.

\section{The Cross-Entropy Wavemap Observable}

The previous section introduced the signal vocabulary for a wave with a single
origin: one center fires, the wavefront spreads, and the red-blue sign flip
at the bond endpoints is the detection signal. Here we extend to the case
where $p$ centers fire independently, each producing its own wavemap. The key
quantities are defined per-center, then combined into an ensemble.

For each center $k$ and site $v$, we evaluate the signed cross-entropy in a
3-cycle window $[t_0 \pm 1]$ centered on the BFS frontier crossing
$t_0(v) = \min\{t : n^k_{\rm base}(v,t) > \varepsilon_{\rm th}\}$:
\begin{equation}
    \mathcal{L}_k(v,\, [t_0 \pm 1]) = -\ln \frac{n^k_{\rm pert}(v,\, t_0(v))}
                                                  {n^k_{\rm base}(v,\, t_0(v))}
\end{equation}
This is a \emph{cross-correlation fixed in space}: both waves are evaluated
at the same site $v$ and the same frontier moment $t_0(v)$, defined by the
baseline. The window $[t_0-1, t_0, t_0+1]$ captures three states ---
before arrival (WHITE), wall-face signal (BLUE or WHITE), and shadow signal (RED) ---
and $t_0(v)$ is the moment of maximum signal and minimum $\chi$-saturation noise.

\begin{figure*}[!t]
\centering
\includegraphics[width=\textwidth]{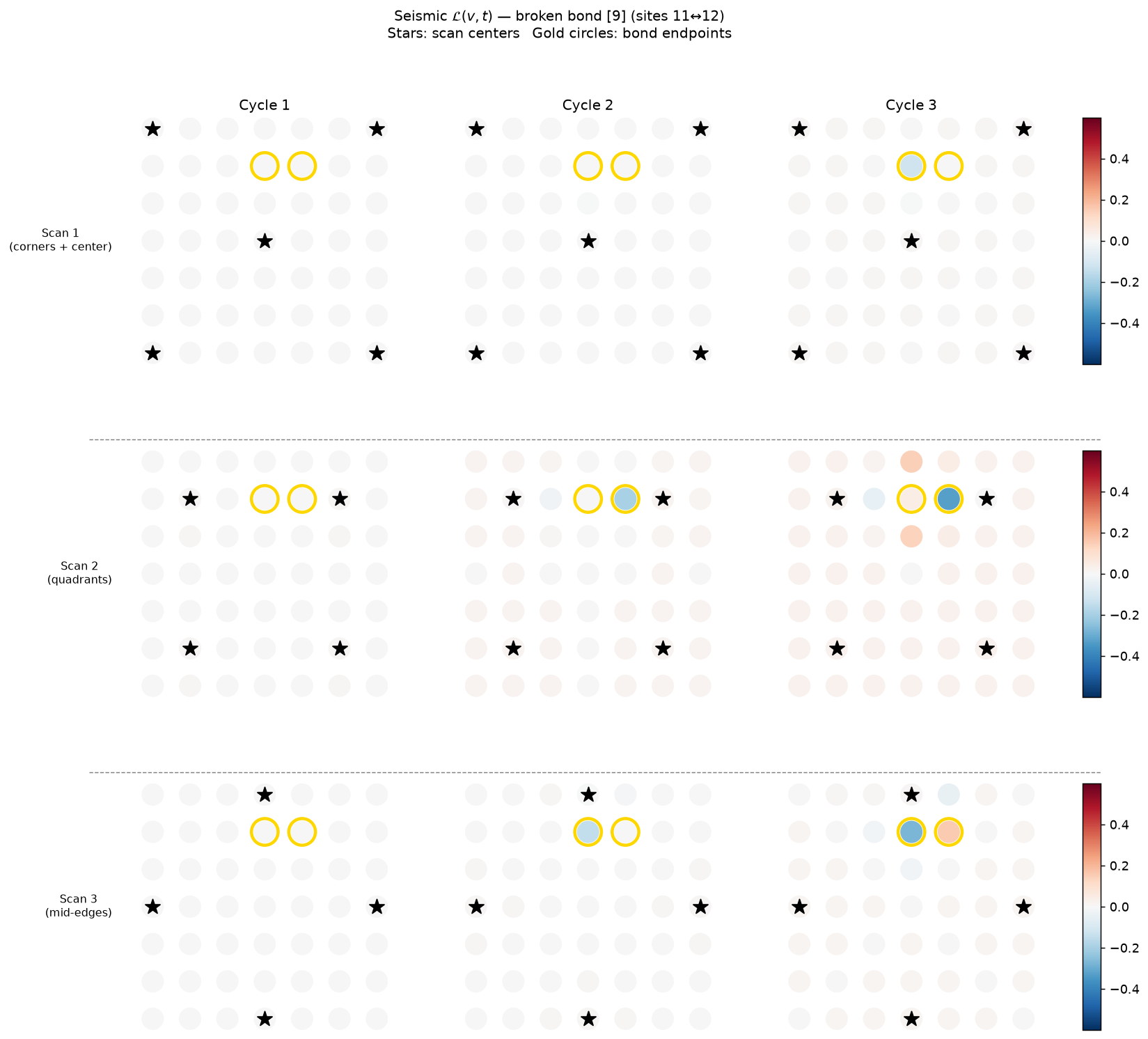}
\caption{Signed cross-entropy $\mathcal{L}(v,t)$ for a single broken bond
(bond 9, sites 11$\leftrightarrow$12, $\delta J=-0.20$),
rect $7\times7$, $\chi=50$, 3 cycles.
Rows show the three seismic scans: corners+center, quadrants, mid-edges.
Columns show cycles 1--3.
\textbf{Cycle 1}: no signal in any scan.
\textbf{Cycle 2}: yellow (BLUE) accumulation visible at the wall-face site.
\textbf{Cycle 3}: red (RED) shadow appears at the downstream site; sign flip
across the bond identifies it in all three scans independently.
Gold circles: bond endpoints. Stars: scan centers.}
\label{fig:ce}
\end{figure*}

Consider a bond $(u, v)$ with $u$ closer to center $k$.
At time $t_0$, the baseline wave has just crossed the bond and
$n_{\rm base}(u, t_0) \approx n_{\rm pert}(u, t_0)$; both waves have reached $u$
with equal amplitude. One cycle later at $t_1 = t_0 + 1$, the weaker bond
slows drainage from $u$: weight accumulates, $n_{\rm pert}(u) > n_{\rm base}(u)$,
so $\mathcal{L}(u) < 0$ (blue, wall face). Simultaneously, in the baseline
weight flows through bond $(u,v)$ and arrives at $v$, while in the perturbed run
less weight crosses: $n_{\rm pert}(v, t_1) < n_{\rm base}(v, t_1)$,
so $\mathcal{L}(v) > 0$ (red, shadow). The temporal order is universal: blue
at the wall face first, red at the shadow one cycle later. The broken bond sits
between the blue and red sites, identified by the sign change (Figure~\ref{fig:ce}).
The contrast $|\mathcal{L}(u) - \mathcal{L}(v)|$ is maximum at $t_1$, providing
the detection signal.

Each center $k$ is initialized and evolved \emph{independently}: a separate
baseline and perturbed simulation pair per center. The signed cross-entropy
$\mathcal{L}_k(v)$ is computed for each center individually, then summed into
an ensemble log-likelihood:
\begin{equation}
    \mathcal{L}_{\rm ens}(v) = \sum_{k=1}^{p} \mathcal{L}_k(v)
\end{equation}
Centers on the same side of the broken bond reinforce each other; centers on
opposite sides contribute opposite signs that partially cancel. Silent centers
(wave never reaches site $v$) contribute zero and do not corrupt the sum.

A more robust alternative to the pointwise $\mathcal{L}(v, t_0)$ accumulates
the signal over all $T$ cycles:
\begin{equation}
    \mathcal{B}(v) = \sum_{t=1}^{T} n_{\rm base}(v,t), \qquad
    \mathcal{N}(v) = \sum_{t=1}^{T} n_{\rm pert}(v,t), \qquad
    \Lambda(v) = \ln \frac{\mathcal{B}(v)}{\mathcal{N}(v)}
\end{equation}
Positive $\Lambda$ identifies red (shadow), negative identifies blue (wall face).
The integral captures both the delay and the sustained amplitude difference: a
broken bond delays the perturbed wave \emph{and} reduces its steady-state weight,
making $\Lambda(v)$ more robust than $\mathcal{L}(v, t_0)$ when the pointwise
signal is weak. In the multi-bond experiments, $\Lambda(v)$ is the primary
detection signal.

\section{The Log-Transfer Vector Field}
\label{sec:logtransfer}

The cross-entropy $\mathcal{L}(v,t) = \ln(n_{\rm base}/n_{\rm pert})$ is a
\emph{scalar field} on sites. We now show that its natural spatio-temporal
gradient is a \emph{vector field} defined on bonds, and that this gradient
field directly reveals the broken bond as a physical singularity.

Since $n(v,t)$ is a scalar field on the graph, its discrete spatial gradient
at site $v$ in the direction of neighbor $u$ is $(\nabla n)_{v\to u}(t) =
n(u,t) - n(v,t)$. Since $n(v,t)$ is a probability, weight flows \emph{multiplicatively}
through bonds: $n(u,t+1) = n(v,t)\cdot T_{v\to u}$, so the natural
gradient is in log-space. The \emph{spatio-temporal gradient} of
$\mathcal{L}$ at directed bond $(v \to u)$ from cycle $t$ to $t+1$ is:
\begin{align}
    (\nabla_{\!\rm st}\,\mathcal{L})_{v \to u}(t)
    &\;=\; \mathcal{L}(u,\,t+1) - \mathcal{L}(v,\,t) \nonumber\\
    &\;=\; \ln\frac{n_{\rm base}(u,t+1)}{n_{\rm base}(v,t)}
         -\ln\frac{n_{\rm pert}(u,t+1)}{n_{\rm pert}(v,t)}
    \;=\; \ln\frac{T^{\rm base}}{T^{\rm pert}}
    \;\equiv\; \delta\!\ln T_{v \to u}(t)
    \label{eq:dlogT}
\end{align}
At each site $v$, collecting this quantity over all neighbors
$u_1, u_2, \ldots$ (up to 4 for a rectangular lattice) gives a
\emph{vector field}:
\begin{equation}
    (\nabla_{\!\rm st}\,\mathcal{L})_v(t)
    \;=\; \bigl[\,(\nabla_{\!\rm st}\,\mathcal{L})_{v \to u_1}(t),\;
                  (\nabla_{\!\rm st}\,\mathcal{L})_{v \to u_2}(t),\;
                  \ldots\,\bigr]
    \label{eq:gradvec}
\end{equation}
one component per bond direction. Each component is a log-likelihood
ratio: positive when the baseline transfers more than the perturbed run
along that bond, negative when the perturbed transfers more.

A broken bond $(u,v)$ is a \emph{singularity} in this gradient field.
At the first wavefront crossing (cycle $t$), the baseline flows through
the bond freely while the perturbed run is impeded:
$\delta\!\ln T_{u \to v}(t) > 0$ (red, baseline transfers more through
the broken bond). One cycle later, weight has accumulated at the wall-face
site $u$ in the perturbed run; $n_{\rm pert}(u)$ grows, making the
reverse ratio $n(v,t+1)/n_{\rm pert}(u,t)$ larger:
$\delta\!\ln T_{v \to u}(t) < 0$ (blue, perturbed transfers more in
reverse because the shadow site $v$ is depleted). The two directed
components of the same bond have \emph{opposite signs}: the gradient
field has a discontinuity exactly at the broken bond.

\begin{figure*}[!t]
\centering
\includegraphics[width=0.8\textwidth]{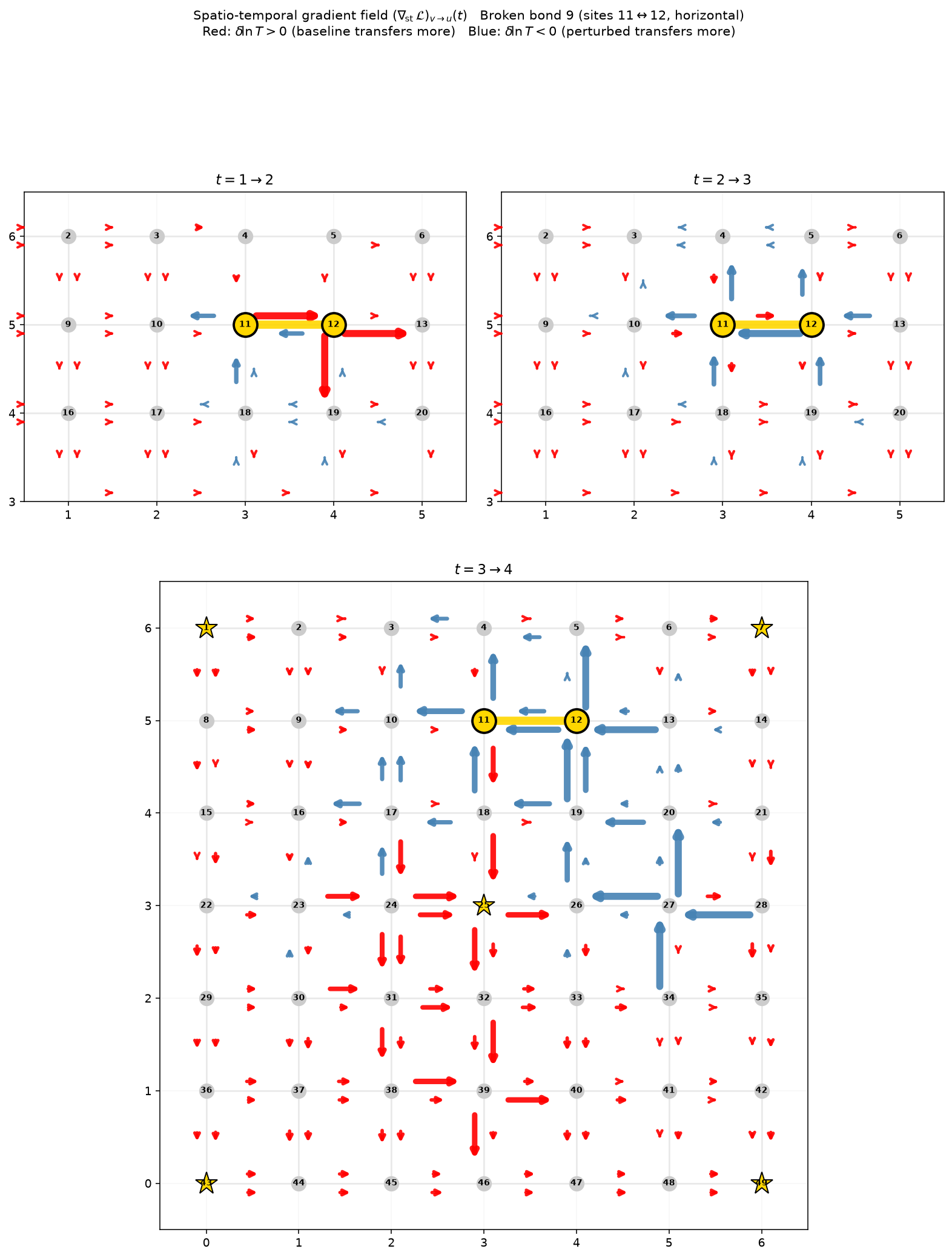}
\caption{$(\nabla_{\!\rm st}\,\mathcal{L})_{v\to u}(t)$,
broken bond 9 (sites 11$\leftrightarrow$12), $\chi=50$.
Red: baseline transfers more. Blue: perturbed transfers more.}
\label{fig:dlogT}
\end{figure*}

By subsequent cycles the perturbation diffracts: the wall-face
accumulation radiates excess weight along alternative paths, enriching
bonds that would normally receive less, while the shadow deepens. The
gradient field develops a characteristic \emph{diffraction pattern}
centered on the broken bond, with two diagonal fringes whose orientation
encodes the bond direction directly (Figure~\ref{fig:dlogT}).

\clearpage

The gradient field $\nabla_{\!\rm st}\,\mathcal{L}$ has several properties
that distinguish it from the scalar $\mathcal{L}(v,t)$. It is defined on
\emph{bonds} rather than sites, so the broken bond appears directly as a
gradient discontinuity rather than through its effect on neighboring sites.
It carries \emph{directional} information: the sign asymmetry between
forward and reverse components identifies which side is the wall face and
which is the shadow, and the diffraction pattern orientation encodes the
bond direction. The normalization by $n(v,t)$ suppresses the dominant-site
bias that affects site-based metrics. Systematic artifacts from the
simulation (Trotter color ordering, BP approximation, or any other
effect present in both runs) cancel in the ratio $T_{\rm base}/T_{\rm pert}$,
whether $n_{\rm base}$ and $n_{\rm pert}$ come from two simulations or
from a simulation and hardware shots.

\section{Seismic Multi-Center Scanning}

Rather than running each center independently, we initialize all $p$ centers
simultaneously in a single seismic scan: the initial operator is
$O(0) = \sum_k Z_{c_k}$, firing all centers at once. At each site $v$,
the first-arrival signal is dominated by the nearest center, automatically
selecting the strongest directional component. The sign flip at the broken bond
endpoints emerges without any explicit path tracing or per-center bookkeeping,
and the cost drops from $2p$ simulations to just 2 (one baseline, one perturbed).

To achieve broad coverage with no geometric blind spots, we use three
complementary center placements on the $7\times7$ lattice: corners and center
($\{1,7,25,43,49\}$), quadrant centers ($\{11,13,37,39\}$), and mid-edge
sites ($\{4,22,28,46\}$). Each placement covers a different region of the
grid; together the three scans ensure every bond is within $d_{\rm max}=2$
hops of at least one center in at least one scan. At $v_{\rm hop}\approx0.45$
hops/cycle, 2 hops requires 5 simulation cycles. The total cost is 6 simulations
(3 baselines + 3 perturbed), run in parallel on 3 GPUs.

Bond scores are computed using the integrated log-ratio $\Lambda(v)$ across
all three scans. For each bond $b$ connecting sites $v_a$ and $v_b$, the
score is the maximum contrast $|\Lambda(v_a) - \Lambda(v_b)|$ over all scans.
The top-$k$ bonds by score are declared broken, where $k$ is the number of
expected faults. On 20 single-bond trials ($\delta J = -0.20$,
two independent datasets), this procedure achieves \textbf{20/20} correct
identifications. On 10 two-bond trials it achieves \textbf{9/10}, with the
single failure corresponding to two bonds sharing an endpoint whose wavefronts
cancel (the BLUE-BLUE-RED interference pattern discussed in Section~\ref{sec:logtransfer}).
\section{Bond Detection}

For each bond $b$ connecting sites $v_a$ and $v_b$ we define a {\em
  bond score}:
\begin{equation}
    S(b) = \mathcal{L}_{\rm ens}(v_a) \times \mathcal{L}_{\rm ens}(v_b)
\end{equation}
A negative product indicates a sign flip across the bond, the causal
signature of the broken bond. Rank bonds by most negative $S(b)$.

Table~\ref{tab:results} shows the individual and ensemble results for
bond 12 (sites $15\\leftrightarrow16$, identity PTM perturbation, $7\times7$ lattice,
$\chi=50$, 5 cycles).

\begin{table}[h]
\centering
\small
\begin{tabular}{lcrrl}
\toprule
Center & Side & $\mathcal{L}(15)$ & $\mathcal{L}(16)$ & Result \\
\midrule
10 & upstream   & $+0.87$ & $-0.02$ & \checkmark CORRECT \\
30 & upstream   & $+0.37$ & $-0.04$ & \checkmark CORRECT \\
25 & downstream & $+0.03$ & $-0.04$ & $\times$ WRONG \\
32 & downstream & $+0.00$ & $-0.03$ & $\times$ WRONG \\
6  & too far    & $0$     & $0$     & silent \\
34 & too far    & $0$     & $0$     & silent \\
\midrule
Ensemble & — & $+1.23$ & $-0.13$ & \checkmark CORRECT \\
\bottomrule
\end{tabular}
\caption{Signed cross-entropy at bond 12 endpoints (sites 15,16).
Ensemble score: $S(12) = -0.164$, next bond $= -0.002$ (82× margin).}
\label{tab:results}
\end{table}

The ensemble identifies bond 12 with an 82-fold margin over the next
candidate, despite only 2 of 6 individual centers giving the correct answer.
The wrong individual answers: centers 25 and 32 both correctly identify
site 16 as a bond endpoint; their error is only in declaring the wrong
adjacent bond. The ensemble product cancels these partial errors.

The gradient field of Section~\ref{sec:logtransfer} provides a complementary
bond score. For each directed bond $(v\to u)$, summing $|\delta\!\ln T_{v\to u}(t)|$
above a threshold over all scans and cycles gives:
\begin{equation}
    \sigma(b) = \sum_{\rm scans} \sum_t
    \bigl(|\delta\!\ln T_{v\to u}(t)| + |\delta\!\ln T_{u\to v}(t)|\bigr)
    \cdot \mathbf{1}[|\delta\!\ln T| > \tau]
    \label{eq:dlogT_score}
\end{equation}
with threshold $\tau=0.01$. The broken bond is the one with the largest
asymmetry between forward and reverse directions. Table~\ref{tab:comparison}
compares the two methods across bond counts.

\begin{table}[h]
\centering
\small
\begin{tabular}{rcc}
\toprule
$k$ bonds & $S(b)$ from $\mathcal{L}$ & $\sigma(b)$ from $\nabla_{\rm st}\mathcal{L}$ \\
\midrule
1  & \textbf{20/20} & \textbf{20/20} \\
2  & \textbf{9/10}  & \textbf{9/10}  \\
4  & 3/5 ALL        & 1/5 ALL        \\
8  & mixed          & avg 6/8        \\
\bottomrule
\end{tabular}
\caption{Detection rate (ALL bonds correct) for the two bond scores.
Both achieve 20/20 single-bond and 9/10 two-bond. They fail on
\emph{different} trials at higher bond counts, making them complementary.}
\label{tab:comparison}
\end{table}

Both methods support an iterative calibration strategy: declare the
top-$k$ broken bonds, recalibrate the hardware, then re-run the scan.
In each pass the already-fixed bonds no longer contaminate the signal,
so the remaining faults surface more clearly. The process stops when
$n_{\rm pert}(v,t) \approx n_{\rm base}(v,t)$ everywhere: a physical
criterion that requires no additional measurement overhead.

\section{Tensor Network Simulation as Quantum Circuit Execution}
\label{sec:tn}

The Trotter simulation underlying our method is not merely an approximation
of a quantum circuit; it \emph{is} the quantum circuit, represented in the
Pauli operator basis rather than the amplitude basis. The identification is exact:

\begin{itemize}
    \item Each \textbf{lattice site} = one qubit
    \item Each \textbf{bond gate} $\exp(-i\theta(XX+YY+ZZ))$ = one two-qubit gate
    \item Each \textbf{color class} (non-overlapping bonds applied in parallel) = one gate layer
    \item Each \textbf{belief propagation synchronization} after a full Trotter cycle = one barrier
    (per-color-class BP, as in the Julia L-MPF implementation~\cite{robertson2026lindblad},
    provides finer sub-cycle resolution matching the quantum circuit barrier structure)
    \item Each \textbf{PAULI\_NL output} = a mid-circuit Pauli weight measurement
\end{itemize}

Both the graph coloring (determining which bonds share qubits) and the quantum
circuit parallelism constraint (which gates can be applied simultaneously) derive
from the lattice topology. The color class ordering is the gate layer ordering.

This identification reveals two distinct timescales. \textit{Physical time}
$t$ counts integer Trotter cycles, $t = n \cdot \varepsilon$; the observables
$C(t)$ and $n(v,t)$ at cycle granularity are physical, independent of color
class ordering to $O(\varepsilon^2)$ and convergent to true quantum dynamics.
\textit{Circuit time} $t_{\rm color}$ has sub-cycle, per-color-class granularity,
a property of the circuit implementation rather than the Hamiltonian. Two Trotter
implementations of the same Hamiltonian with different color orderings agree on
$C(t)$ to first order in $\varepsilon$ but differ by $O(\varepsilon^2)$ Trotter
errors; their $n(v, t_{\rm color})$ tomography fingerprints differ already at the
first sub-cycle. The BP synchronization granularity determines which timescale is
accessible: per-cycle BP (as used here) loses sub-step information; per-color BP
(Julia L-MPF~\cite{robertson2026lindblad}) respects the causal barrier structure
of the quantum circuit.

A perturbed bond (coupling $J \neq 1$) is a \emph{miscalibrated gate}: the
two-qubit rotation angle $\theta = J \cdot \varepsilon$ differs from the nominal
value. The Pauli transfer matrix for $\exp(-i\theta(XX+YY+ZZ))$ has diagonal
entries $\cos(\theta)$, which for any $J \neq 1$ gives $\cos(J\varepsilon) \neq \cos(\varepsilon)$.
Since $|\cos(\theta)| \leq 1$ for all $\theta$ (unitary gate), any coupling
deviation attenuates the wave relative to baseline, guaranteeing $n_{\rm pert}(v^*) < n_{\rm base}(v^*)$
at the shadow endpoint $v^*$, so $\mathcal{L}(v^*) > 0$ (red) regardless of
the sign of $\delta J = J - 1$. The detection method is robust: the red site
is always the shadow endpoint, never inverted. The gradient field
$\nabla_{\!\rm st}\,\mathcal{L}$ of Section~\ref{sec:logtransfer} reveals
the complementary effect: while the bond attenuates the wave on the shadow
side, it simultaneously causes accumulation on the wall-face side and,
by subsequent cycles, amplification along rerouted paths that would
otherwise receive less weight.

Belief propagation is exact on trees: messages from disjoint subtrees
are independent and the fixed point is unique. The seismic initialization
fires multiple centers simultaneously, creating wavefronts that meet and
interfere inside the graph. This interference violates the tree assumption
structurally: circular message dependencies prevent convergence regardless
of iteration count. It is a property of the multi-center initialization,
not of the solver.

The detection observable $\mathcal{L}(v,t) = \ln(n_{\rm base}/n_{\rm pert})$
is a ratio of two runs that share the same centers, the same graph, and
the same BP schedule. Whatever interference pattern the multi-center
initialization creates is therefore structurally identical in both runs.
It is not an approximation error to be corrected; it is a shared
background that cancels exactly in the ratio. Only the bond-specific
perturbation survives. This was verified empirically: running with up to
30,000 damped BP iterations (which achieve convergence for small center
sets) gives PAULI\_NL values identical to standard 100-iteration BP.
The residual BP non-convergence leaves $n(v,t)$ unchanged.

A complete theoretical characterization of multi-center BP on loopy graphs
remains open and will ultimately be resolved by hardware experiments where
$\hat{n}(v,t)$ is measured directly, with no BP involved.

This frames our tomography as \emph{quantum ATPG} (Automatic Test Pattern
Generation)~\cite{roth1966diagnosis}, the classical circuit testing methodology
generalized to quantum gates. In classical ATPG: fault activation (excite the
stuck-at gate), path sensitization (find an observable path from fault to output),
backward implication (infer the faulty gate). The quantum analog is:
perturbation (change $J$), wavefront propagation (operator spreading from
center to shadow site), and cross-entropy inversion (red site + center
position $\to$ bond identity).

\section{Causal Structure, Hardware Path, and Detection Algorithm}

The observable $n(v,t)$ is accessible on real quantum hardware via local
Pauli measurements after ZNE gate-folding circuits~\cite{giurgica2020digital}.
The cross-entropy $\mathcal{L}(v)$ requires only the ratio of two measurements,
with no absolute calibration. The center placement is determined by the graph
topology alone, independent of the Hamiltonian.

Our method exploits the Lieb-Robinson bound: information cannot propagate
faster than $v_{\rm hop}$ hops/cycle. A broken bond creates a causal anomaly;
the wavefront on one side cannot inform the other. The first-arrival time
$t_{\rm first}(v)$ is observed directly, without knowing $v_{\rm hop}$ in advance.

The seismic 3-scan cost is fixed at 6 simulations regardless of how many bonds
are broken. On the $7\times7$ lattice ($\chi=50$, 5 cycles, AMD MI100 32\,GB GPU),
each scan pair runs in 272\,s using 11\% VRAM with 2 GPUs in parallel.

When two adjacent broken bonds share an endpoint, say bonds $(i,j)$ and
$(j,k)$ with shared site $j$, the cross-entropy field shows a three-site
pattern: site $i$ appears red (shadow of bond $(i,j)$), site $k$ appears
blue (wall face of bond $(j,k)$), and the shared site $j$ appears white.
At $j$, the red contribution from one broken bond and the blue contribution
from the other cancel, producing an \emph{interference node}. The white
fringe is not a detection failure; it is an interference pattern arising
from opposite-side contributions canceling at the shared node.
The current algorithm detects such pairs via a triplet scorer
$S_{\rm wall} = \mathcal{L}(v_i) \times \mathcal{L}(v_k)$, skipping the
white shared endpoint and reading the outer sites directly.

The seismic and aperture methods are naturally complementary and suggest
a two-phase adaptive protocol. In \textbf{Phase 1} (seismic), all $p$ centers
fire simultaneously: one quantum experiment, minimum cost, producing a map of
RED sites that identifies the region of interest. In \textbf{Phase 2} (aperture),
the RED sites from Phase 1 guide the selection of targeted observation centers
geometrically positioned to discriminate between candidate bonds adjacent to
those sites. Only $k \ll p$ centers need to be run, each independently.

This mirrors standard practice in other imaging domains: a wide-field scout scan
(seismic) followed by targeted high-resolution acquisition (aperture). On quantum
hardware, Phase 1 maps directly to a single quantum experiment; Phase 2 selects
the minimum number of additional experiments to confirm the bond identity.

The seismic phase requires only one circuit (all centers in a single
initialization) and one set of hardware shots for the perturbed device. The baseline $n_{\rm base}(v,t)$ is provided by
simulation and reused across all candidate perturbations at no additional
hardware cost. The aperture phase then tests the hypothesis with directional
projections; perturbation signals that survive across independent projections
with geometry-consistent timing are physical, while chi-truncation artifacts
and shot noise are uncorrelated across runs and average out.

The following describes wavefront triangulation across scans.
When multiple independent scans are available, the temporal order in which
each scan detects the broken bond carries geometric information beyond the
sign flip alone. This enables a triangulation strategy analogous to GPS:
each scan provides a time-of-arrival constraint that localizes the bond.

Formally, scan $s$ with nearest center $c_s$ detects the broken bond at
cycle $t_s^*$, the first cycle where $|\mathcal{L}_k(v)| > \varepsilon_{\rm th}$
for some bond endpoint $v$. This constrains the bond to lie within
$d_s = t_s^* \times v_{\rm hop}$ hops of $c_s$: a ball $\mathcal{D}(c_s, d_s)$
on the graph. Across $M$ scans, the bond lies in the intersection:
\begin{equation}
    b^* \in \bigcap_{s=1}^{M} \mathcal{D}(c_s,\, t_s^* \times v_{\rm hop})
\end{equation}
where $\mathcal{D}(c, r)$ is the set of bonds within crossing distance $r$ of center $c$.

The sign of the earliest detection (which endpoint is RED vs BLUE) further
constrains the direction: the center $c_s$ is on the BLUE (wall-face) side
of the broken bond. Subsequent scans confirm or deny: a scan detecting the
\emph{opposite} sign at the same endpoints confirms (center is on the other
side of the same bond); a scan detecting different endpoints shifts the
candidate set. Each confirmation tightens the intersection; each denial
eliminates candidates.

In practice on the $7\times7$ lattice with 3 scans (corners+center,
quadrants, mid-edges): scan 2 detects bond 9 (11$\leftrightarrow$12)
at cycle 1 (center 13, 1 hop away), scan 3 confirms with opposite sign at
cycle 2 (center 4, 2 hops away), scan 1 provides weak confirmation at
cycle 2 from the far corners. The intersection of the three balls gives
a small candidate set containing the true bond.

\subsection*{BFS-guided fault detection algorithm}

The detection method reduces to a graph-theoretic algorithm that operates
entirely on the observable $n(v,t)$, without reference to the Hamiltonian,
coupling values, or simulation parameters.
Let $G = (V, E)$ be the coupling graph, $\{c_s\}$ the observation centers,
and $\varepsilon_{\rm th}$ the detection threshold.

\begin{enumerate}
    \item For each center $c_s$ and each site $v \in V$, record the
    baseline arrival time $t^*_{\rm base}(v) = \min\{t : n_{\rm base}(v,t) > \varepsilon_{\rm th}\}$
    and perturbed arrival time $t^*_{\rm pert}(v) = \min\{t : n_{\rm pert}(v,t) > \varepsilon_{\rm th}\}$.
    These are the actual BFS frontier crossings, not predicted hop distances;
    $t^*_{\rm pert}(v)$ may exceed $t^*_{\rm base}(v)$ if a broken bond delays the perturbed wave.

    \item Both the simulation and the quantum circuit assume an execution order
    among independent bonds, determined by Trotter color class. We materialize
    this order for causal determination: assign each site $v$ the pair $(h(v), c(e_v))$
    where $h(v)$ is its BFS hop depth and $c(e_v) \in \{0,1,2,3\}$ is the color
    class of the bond used to reach it. The total order $(h, c)$ gives exact
    sub-cycle causality: $(h_u, c_u) < (h_v, c_v)$ means site $u$ is strictly
    upstream of $v$.

    \item For a bond $(u, v)$ with $u$ upstream of $v$, compute the bond
    transfer coefficient:
    \begin{equation}
        T = \frac{n(v,\, t^*_{\rm base}(v))}{n(u,\, t^*_{\rm base}(u))}
    \end{equation}
    When $n_{\rm base}(u) \gg \varepsilon_{\rm th}$, both waves share the same
    source amplitude at the BFS frontier and
    $\mathcal{L}(v) \approx \ln(T_{\rm base}/T_{\rm pert})$:
    the signed cross-entropy directly measures the relative change in bond transfer.
    A broken bond reduces $T_{\rm pert} < T_{\rm base}$, giving $\mathcal{L}(v) > 0$
    (red shadow) and $\mathcal{L}(u) < 0$ (blue wall face).

    \item For each site $v$ with $|\mathcal{L}(v)| > \varepsilon_{\rm th}$, examine
    neighbors $u$ with $(h_u, c_u) < (h_v, c_v)$. Declare bond $(u,v)$ as a candidate
    if $\mathcal{L}(u)$ and $\mathcal{L}(v)$ have opposite signs. Score each candidate
    by $|\mathcal{L}(u) - \mathcal{L}(v)| \cdot \delta / t^*$ where $\delta = 2$ if
    $t^*_{\rm pert}(v) > t^*_{\rm base}(v)$, else $1$, and
    $t^* = \min(t^*_{\rm base}(u), t^*_{\rm base}(v))$.

    \item Repeat steps 1--4 for each scan independently. Rank all candidate bonds
    by total score across scans and declare the top-$k$ as broken.
\end{enumerate}

From our experiments we have identified three recurrent interference signatures,
which we name here as a working vocabulary based on observed cases. Each is a
consequence of the BFS frontier geometry and the Trotter color class ordering,
with no quantum-specific reasoning required. We expect further patterns to emerge
as the geometry of faults and center placements varies; the vocabulary below
reflects our current observations, not an exhaustive classification.
\textbf{RED-BLUE} is the clean single-bond signature: center on one side,
sign flip across the bond. \textbf{BLUE-WHITE-BLUE} arises when a bond is
sandwiched between two equidistant centers: the shared site is white (interference
node) and the outer sites are blue on both sides, making both adjacent bonds
identifiable. \textbf{BLUE-BLUE-RED} occurs when two orthogonal broken bonds
share an endpoint: the shared site accumulates from two directions, the outer
RED identifies one bond, and the adjacent BLUEs identify the second.

\section{Measurability and Detection Thresholds}

The detection method requires $n(v,t)$ to be measurable both from the
simulation fidelity standpoint and from the hardware shot budget.
Two independent conditions must hold simultaneously.

Recall that $n(v,t) \in [0,1]$ is a probability: the fraction of Pauli basis
operators that have non-identity support at site $v$ after $t$ Trotter cycles
(Section~\ref{sec:vocab}). On real hardware the circuit is executed $N$ times.
A barrier after each Trotter cycle allows a mid-circuit local Pauli weight
measurement at site $v$; each shot returns either the identity at $v$
or a non-identity Pauli ($X$, $Y$, or $Z$). The three non-identity outcomes
are not distinguished: only the binary question ``identity or not'' is asked,
and the non-identity count $k$ accumulates as a success count. This binary
nature of each shot makes the Binomial model natural: $k \sim
\mathrm{Binomial}(N, n(v,t))$, where the physical probability $n(v,t)$ is
the Bernoulli success probability for each of the $N$ independent shots.
The empirical estimate $\hat{n}(v,t) = k/N$ is then an unbiased estimator
of the true $n(v,t)$.

The estimator has variance $\mathrm{Var}(\hat{n}) = n(1-n)/N$, a direct
consequence of the Binomial distribution. At the frontier, $n(v,t) \ll 1$:
the total Pauli weight is conserved ($\sum_v n(v,t) = 1$) and spreads over
a growing shell of sites, so each individual $n(v,t)$ decreases with distance
from the center. In this small-$n$ regime $1-n \approx 1$ and the variance
simplifies to $n/N$, giving a shot noise floor $\sigma(n) \approx \sqrt{n/N}$.
Note also that only sites where $n(v,t) > \varepsilon_{\rm th}$ contribute
to the detection (Section~\ref{sec:vocab}), so the relevant regime is
$\varepsilon_{\rm th} \leq n(v,t) \ll 1$.
The baseline $n_{\rm base}(v,t)$ is provided by simulation and is exact; only
the perturbed run is measured on hardware. The condition is that
$\hat{n}_{\rm pert}$ is a reliable estimate of the true $n_{\rm pert}(v,t)$,
so that the comparison with the theoretical $n_{\rm base}(v,t)$ is meaningful.
Reliability requires the standard deviation of the estimator $\sigma(\hat{n})
= \sqrt{n/N}$ to be much smaller than the quantity being estimated: $\sigma(\hat{n}) \ll n$.
As $N$ grows, $\sigma(\hat{n})$ shrinks and the estimate converges to the true value.
The minimum observable $n$ is set by the detection threshold $\varepsilon_{\rm th}$.
For $\hat{n}$ to be a reliable estimate at $n = \varepsilon_{\rm th}$ we require
$\sigma(\hat{n}) \leq \varepsilon_{\rm th}$, i.e.,
$\sqrt{\varepsilon_{\rm th}/N} \leq \varepsilon_{\rm th}$, which gives:
\begin{equation}
    N \geq \frac{1}{\varepsilon_{\rm th}}
\end{equation}
For $\varepsilon_{\rm th} = 10^{-3}$, at least $10^3$ shots are required.
Table~\ref{tab:shots} shows the statistical floor as a function of shot budget.

\begin{table}[h]
\centering
\small
\begin{tabular}{rrl}
\toprule
$N$ shots & $\sigma(\hat{n})$ at $n=\varepsilon_{\rm th}$ & Status \\
\midrule
$10^2$   & $3 \times 10^{-3}$ & noise exceeds threshold \\
$10^3$   & $1 \times 10^{-3}$ & noise matches threshold (minimum) \\
$10^4$   & $3 \times 10^{-4}$ & reliable estimate \\
$10^6$   & $3 \times 10^{-5}$ & well into reliable regime \\
\bottomrule
\end{tabular}
\caption{Estimator noise $\sigma(\hat{n}) = \sqrt{\varepsilon_{\rm th}/N}$
at the detection threshold $\varepsilon_{\rm th} = 10^{-3}$.
Reliable detection requires $N \geq 1/\varepsilon_{\rm th} = 10^3$.}
\label{tab:shots}
\end{table}

The tensor network simulation introduces two distinct noise sources that set a
floor on trustworthy $n(v,t)$ values. First, truncation: bond singular values
below the cutoff $\delta_{\rm cut}$ are set to zero, so any $n(v,t)$ arising
from truncated components is numerical noise, not physical operator spreading.
Second, belief propagation is not strictly causal: messages propagate globally
in each sweep, producing tiny non-zero $n(v,t) \sim 10^{-4}$ at sites outside
the Lieb-Robinson lightcone. In a real quantum circuit these sites would never
be activated (only gates within the lightcone fire), so this leakage is a
simulation artifact with no hardware counterpart. Both effects are below
$\delta_{\rm cut} = 10^{-3}$, which we therefore adopt as the simulation floor.

Other simulators with tighter cutoffs could in principle trust smaller $n(v,t)$
values and use a lower $\varepsilon_{\rm th}$. However, a smaller threshold
demands proportionally more shots at the frontier: to estimate $n \approx
\delta_{\rm cut}$ reliably one needs $N \geq 1/\delta_{\rm cut}$ shots.
A simulation floor of $10^{-6}$ would require $10^6$ shots merely to reach
reliable estimates at the threshold, a cost that must be weighed against
the physical benefit of the finer resolution.

The two conditions converge at the same threshold $\varepsilon_{\rm th} =
\delta_{\rm cut} = 10^{-3}$: the simulation floor sets the minimum trustworthy
signal, and from Table~\ref{tab:shots} reliable hardware estimates at that
level require $N \geq 1/\varepsilon_{\rm th} = 10^3$ shots. Both conditions
are satisfied simultaneously by:
\begin{equation}
    \varepsilon_{\rm th} \geq \delta_{\rm cut}, \qquad N \geq \frac{1}{\varepsilon_{\rm th}}
\end{equation}
A tighter simulation cutoff $\delta_{\rm cut} < 10^{-3}$ enables a smaller
$\varepsilon_{\rm th}$ and finer resolution, but demands proportionally more
shots at the frontier to maintain reliable estimates.

From Section~\ref{sec:tn}, the bond gate PTM has diagonal entries $\cos(J\varepsilon)$.
A coupling deviation $\delta J$ shifts the rotation angle and attenuates the diagonal
entry by $-\sin(J\varepsilon)\cdot\varepsilon\,\delta J$ to first order. At $J=1$
this means the perturbed weight at site $v$ is simply the baseline attenuated:
\begin{equation}
    n_{\rm pert}(v,t) \approx n_{\rm base}(v,t)\,\bigl[1 - \varepsilon\sin(\varepsilon)|\delta J|\bigr]
\end{equation}
so the signal $\delta n = n_{\rm base} - n_{\rm pert} \approx n_{\rm base}(v,t)
\cdot \varepsilon \cdot \sin(\varepsilon) \cdot |\delta J|$ is proportional to the
baseline weight and scales linearly with the coupling deviation. The attenuation
factor $\varepsilon\sin(\varepsilon)|\delta J|$ depends only on the Trotter step
and the perturbation magnitude, not on the site.

Note that this is a stronger requirement than simply estimating $\hat{n}_{\rm pert}$
reliably: we need the signal $\delta n$ to be measurably larger than the estimator
noise $\sigma(\hat{n}) \approx \sqrt{n_{\rm base}/N}$. Setting $\delta n \geq
\sigma(\hat{n})$ gives the minimum detectable coupling deviation.
We emphasize that the present work performs \emph{detection}: identifying the location and existence of a broken bond from wavemap patterns, not
\emph{estimation} of $\delta J$ from the numerical values of $n(v,t)$.
Quantitative estimation of the coupling deviation from the measured $\hat{n}_{\rm pert}$
is a harder and separate problem, left for future work.
\begin{equation}
    |\delta J|_{\rm min} \approx \frac{\varepsilon_{\rm th}}
        {\varepsilon \cdot \sin(\varepsilon) \cdot \sqrt{N \cdot n_{\rm base}(v,t)}}
\end{equation}
For $\varepsilon = 0.10$ and $n_{\rm base} \approx 10^{-2}$ (genuine wavefront
at first arrival), the detection threshold scales as $|\delta J|_{\rm min}
\propto 1/\sqrt{N}$: at $N=10^6$ shots $|\delta J|_{\rm min} \approx 1.0$,
at $N=10^7$ it drops to $0.32$. To detect the experimental coupling
degradation $|\delta J| = 0.20$ used throughout this work requires
$N \approx 2.5 \times 10^7$ shots. This is independent of the minimum
shot budget for reliable estimation ($N \geq 10^3$): the two conditions
address different questions : can we measure $n(v,t)$ at all, versus can
we resolve the specific coupling deviation of interest.

\section{Conclusion}

A broken bond leaves a causal signature in the operator wavefront: a
red-blue sign flip in $\mathcal{L}(v,t)$ at its two endpoints, identifiable
from the first wavefront crossing and robust across all center placements.
Its spatio-temporal gradient $\nabla_{\!\rm st}\,\mathcal{L}$ reveals the same
bond as a physical singularity: a gradient discontinuity with a characteristic
diffraction pattern whose orientation encodes the bond direction.
Three seismic scans, each firing a different set of centers simultaneously,
cover all 84 bonds of the $7\times7$ lattice in 6 simulations. Both detection
methods achieve 20/20 single-bond and 9/10 two-bond identification, with
failures on complementary trials. The per-bond detection rate remains above
90\% for up to 8 simultaneous faults in a single pass. An iterative protocol
(declare, recalibrate, re-scan) progressively resolves multi-bond cases,
with $n_{\rm pert}(v,t) \approx n_{\rm base}(v,t)$ as the natural stopping criterion.

The measurability analysis shows that $N \geq 10^3$ shots suffice for reliable
estimation of $n(v,t)$ at the detection threshold; resolving the specific
coupling deviation $|\delta J| = 0.20$ requires $N \sim 2.5 \times 10^7$ shots,
a stronger condition that governs hardware deployment.

The method is a direct sequitur to the Lindblad MPF framework. Both use the
same observable $n(v,t)$, the same circuits, and the same hardware protocol;
MPF infers noise from a known Hamiltonian, while this work infers coupling
geometry from a known noise baseline. Two inferences from one measurement,
in opposite directions.

\bibliographystyle{plain}
\bibliography{tomography}

@inproceedings{hong1981pebble,
  author    = {Hong, Jia-Wei and Kung, H. T.},
  title     = {{I/O} Complexity: The Red-Blue Pebble Game},
  booktitle = {Proceedings of the Thirteenth Annual ACM Symposium on Theory of Computing (STOC '81)},
  pages     = {326--333},
  year      = {1981},
  doi       = {10.1145/800076.802486}
}

@article{giurgica2020digital,
  author    = {Giurgica-Tiron, Tudor and Hindy, Yousef and LaRose, Ryan
               and Mari, Andrea and Zeng, William J.},
  title     = {Digital zero noise extrapolation for quantum error mitigation},
  journal   = {2020 IEEE International Conference on Quantum Computing and Engineering (QCE)},
  pages     = {306--316},
  year      = {2020},
  doi       = {10.1109/QCE49297.2020.00045}
}

@misc{robertson2026lindblad,
  author        = {Robertson, Andrew and others},
  title         = {Lindblad Multiproduct Formulas},
  year          = {2026},
  note          = {Manuscript}
}

@article{roth1966diagnosis,
  author  = {Roth, J. P.},
  title   = {Diagnosis of Automata Failures: A Calculus and a Method},
  journal = {IBM Journal of Research and Development},
  volume  = {10},
  number  = {4},
  pages   = {278--291},
  year    = {1966},
  doi     = {10.1147/rd.104.0278}
}

\appendix

\section{Aperture Synthesis with Voronoi Center Placement}
\label{app:voronoi}

An alternative to the seismic approach runs each center independently, in
direct analogy with aperture synthesis in radio interferometry. Each center
$k$ at a different graph position plays the role of one antenna at a different
baseline; $\mathcal{L}_k(v)$ encodes which side of the broken bond site $v$
is on (sign) and the signal strength (magnitude).

Centers are placed with the constraint that no two share the same row or
column in the lattice. This guarantees every bond has a strictly unique nearest
center (no Voronoi ties): the perpendicular bisector between any two centers
never passes through a bond, so one center always dominates the signal at every
bond. For $p=7$ on the $7\times7$ lattice (one center per row, one per column,
a permutation placement), exhaustive search over permutations minimizing
$\max_b T_b$ gives centers $\{3,13,15,25,35,37,47\}$ with zero ties across
all 84 bonds, maximum $T_b = 5$ cycles, and balanced Voronoi regions of
10--16 bonds each.

Ensemble summation $\mathcal{L}_{\rm ens}(v) = \sum_k \mathcal{L}_k(v)$ fails
for bonds sandwiched between centers on opposite sides, where RED and BLUE
contributions partially cancel. A \emph{winner-takes-all} rule avoids this:
for each center $k$ compute the signed contrast across every bond,
\begin{equation}
    C_k(b) = \bigl|\mathcal{L}_k^{\rm signed}(v_a) - \mathcal{L}_k^{\rm signed}(v_b)\bigr|
\end{equation}
and declare the globally winning center-bond pair
$(k^*, b^*) = \argmax_{k,b} C_k(b)$.
The winning center is necessarily on the BLUE side of the broken bond; each
center is a fully independent observer and the one with the strongest signal wins.
On 10 trials ($\delta J = -0.20$), the adaptive seismic-then-aperture protocol
achieves \textbf{10/10} correct identifications: 7 bonds identified by seismic
alone (2 simulations), 3 by aperture (14 simulations), at an average cost of
$2 + 0.3 \times 14 \approx 6$ simulations per bond.

The seismic 3-scan approach in Section~6 supersedes this method for the
multi-bond case: it achieves the same single-bond accuracy at lower cost,
and extends naturally to multiple simultaneous faults without requiring
the no-shared-row-or-column placement constraint.

\end{document}